# SonoTransformers: Ultrafast Transformable Wireless Microscale Machines

Zhiyuan Zhang, Zhan Shi, and Daniel Ahmed*

Acoustic Robotics Systems Lab, Institute of Robotics and Intelligent Systems, Department of Mechanical and Process Engineering, ETH Zurich; Rüschlikon, CH-8803, Switzerland

*Corresponding author: dahmed@ethz.ch

**Abstract**

Shape transformation, a key mechanism for organismal survival and adaptation, has gained importance across fields as diverse as electronics and medicine. However, designing and controlling microscale shape-shifting materials remains a fundamental challenge in various actuation modalities. Here, we introduce "SonoTransformer," an acoustically activated micromachine that delivers ultrafast transformability using preprogrammed soft hinges. These hinges concentrate energy through intensified oscillation and provide the necessary torque for transformation. We have created new machine designs to predetermine the folding state, enabling tailoring and milliseconds transformation. Additionally, we have shown selective transformation by adjusting acoustic power, realizing high degrees of control and functional versatility. Our findings open new research avenues in acoustics, physics, and soft matter, offering new design paradigms and development opportunities in robotics, metamaterials, adaptive optics, and microtechnology.

**Main**

Numerous organisms large and small shift their shapes in order to effectively survive and thrive in complex natural environments. For instance, armadillos curl themselves into a ball to protect themselves from danger, while frilled lizards spread their umbrella-like collar membranes in threat and mating displays. At the microscopic level, unicellular organisms like amoebae extend pseudopods for locomotion and feeding, while a vorticella can contract its spasmoneme within milliseconds [1,2] (**Fig. 1a**). Such shape transformations are facilitated by soft structures, which are characterized by adaptability, flexibility, conformability, energy efficiency, and safety. Drawing inspiration from nature's solutions can serve to inspire innovative artificial systems capable of shape-shifting [3–7], the potential applications of which span fields as diverse as robotics, wearable devices, materials science, and bioengineering.

Shape transformation in artificial systems has been implemented by means of several actuation methods, including mechanical pushes [8,9], pressurized fluids [10–12], shape memory polymers and alloys [13–15], liquid crystal elastomers [16,17], and conductive polymers [18–20]. However, these systems are typically large, often on the scale of centimetres, which precludes their use in miniaturized machines; moreover, they are difficult to scale down due to limitations imposed by materials and manufacturing processes. To date, the design, development, and control of shape-shifting materials at the microscale pose significant challenges. Researchers have utilized a range of external stimuli and fields, including chemical gradients [21–24], light [25–28], heat [29–31], humidity [32], electric and magnetic fields [33–38], and more. These pioneering studies also have limitations; for example, as actuation methods they are often slow, resulting in lower deformation speeds. In addition, sophisticated pre-programming and fabrication processes are needed to produce micromachines compatible with these methods, such as predesigned magnetic moments, multiple layers, and multistep processes. The methods also require special working conditions, such as a particular pH, temperature gradient, or high voltage, and rely on complex material properties such as conductivity and photosensitivity. Recent advancements in electromagnetic actuation have produced fast shape morphing in mechanical two-dimensional (2D) surfaces, but this approach requires tethering, a bulky apparatus, and multiple power sources [39,40]. Throughout all this research into actuation methods for shape transformation, acoustics have received little attention. While the deformation of hinges has previously been utilized in folding mechanisms actuated by various external stimuli [4], the acoustic response associated with a soft hinge remains unexplored. Acoustics presents an exciting alternative that can generate large forces, deliver ultrafast responses, and eliminate the need for complex manipulation systems [41–50].

In this study, we demonstrate acoustically activated ultrafast transformable micromachines that can change their shapes in milliseconds upon the application of an acoustic field and rebound to their original state quickly when the field is deactivated, as illustrated in **Fig. 1b**. These micromachines comprise multiple geometric interconnected acoustic-deformable microbeams, which are serially constructed of rigid links and soft hinges with different stiffnesses. When exposed to an acoustic field, the soft hinge oscillates and induces shape-shifting of the entire microbeam within milliseconds. The force and torque necessary for this shifting are provided by the asymmetrical acoustic radiation force and streaming caused by the oscillation of the soft hinge. We demonstrate the remarkable capacity of acoustic activation to trigger auxetic, isomeric, and selective shape transformation of micromachines depending on their geometric parameters. These results offer new design paradigms for a wide range of applications in soft robotics, flexible electronics, intelligent metamaterials, adaptive optics, multifunctional acoustofluidics, advanced micro-medicines, and beyond.

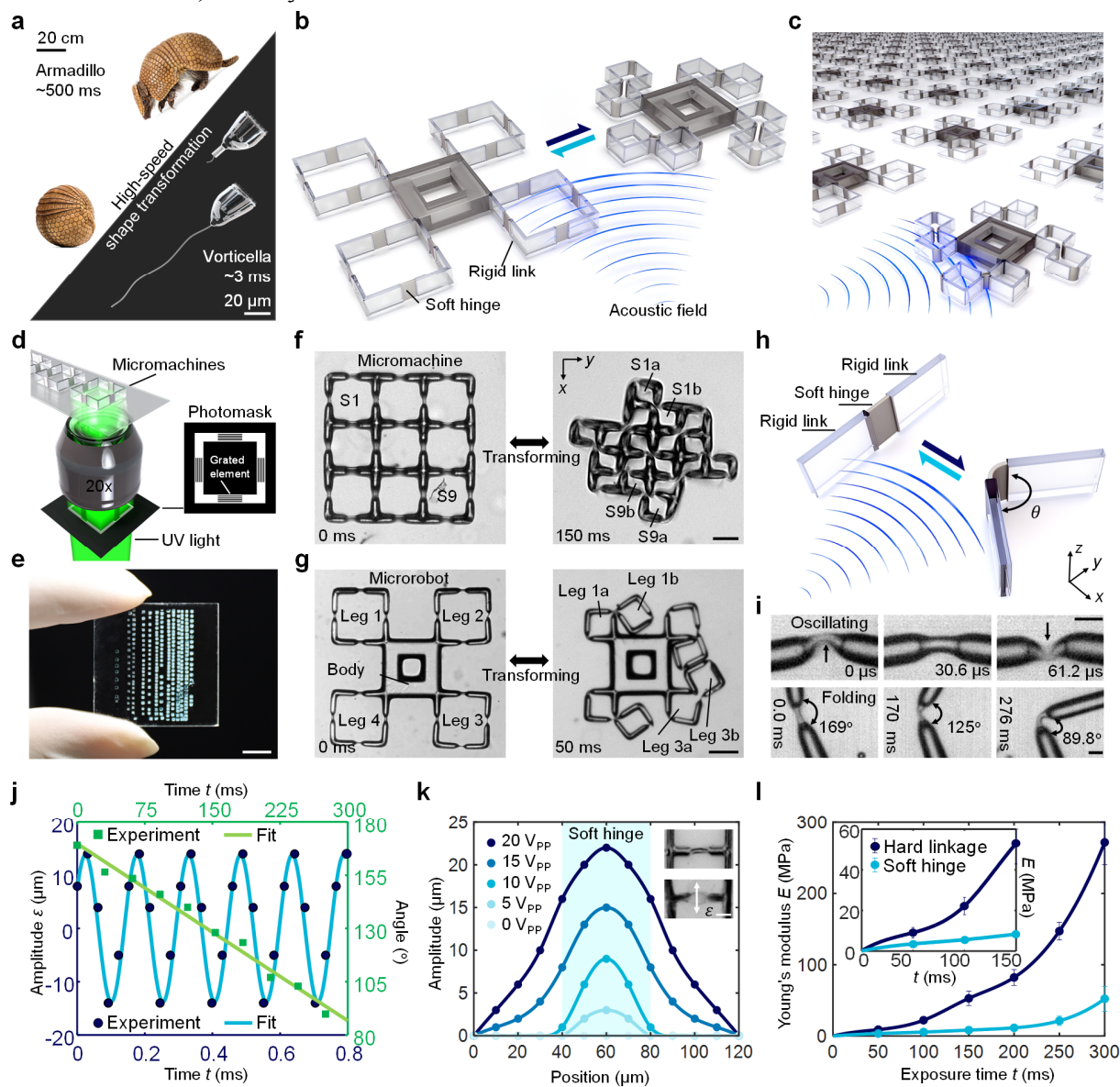

**Fig. 1| Ultrafast acoustically actuated shape transformation. a**, Ultrafast shape transformation occurs across different scales in nature represented by an armadillo and a vorticella. **b**, Schematic of an untethered soft micromachine and its high-speed, reversible shape transformation in an acoustic field. **c**, **d**, The schematic of high throughput micromachine array and photopolymerization setup. Inset shows the photomask with blank and grated elements. **e**, An image shows hundreds of soft micromachines fabricated on a chip. Scale bar, 5 mm. **f**, An untethered microscale auxetic contracts in both length and width from its original shape to a fully folded shape in 150 ms. Scale bar, 100 µm. **g**, A quadrupedal microrobot folds its outer four-square legs in 50 ms. Scale bar, 100 µm. **h**, Illustration of the folding mechanism of an acoustic-deformable microbeam composed of a soft hinge positioned between two rigid links. **i**, image sequences demonstrate the oscillation of the soft hinge and folding of the entire microbeam captured by a high-speed camera. Scale bar, 20 µm. **j**, Plots of the sinusoidal oscillation of the soft

hinge and linear folding of the entire microbeam against time. **k**, Plots of the oscillation amplitude along the acoustic-deformable microbeam at different positions according to the acoustic excitation voltage. Inset shows the oscillation of the clamped-clamped microbeam. Scale bar, 50 μm. **l**, Plots of stiffnesses of the soft hinge and rigid link according to the duration of UV exposure. Inset shows a zoomed-in subplot of the 0-150 ms exposure range.

**Design and characterization of transformable microscale machines**

We fabricated prototypes of the acoustic-transformable micromachines using a custom-built ultraviolet (UV) photopolymerization technique (as shown in **Fig. 1, c** and **d**), which utilizes a photomask and a photosensitive hydrogel mixture on a glass substrate. The designed patterns of the micromachines are projected into the hydrogel mixture as the UV light passes through the photomask, resulting in the polymerization of the hydrogel mixture. To achieve varying stiffnesses and thicknesses in the micromachines, we designed the photomask with blank and grated elements that alter the local exposure intensity. Blank elements result in full polymerization, producing a relatively rigid region, while grated elements enable partial polymerization, resulting in a soft and flexible region (see Supplementary Fig. S1). Here, we designed and fabricated micromachines with varying hinges that comprise 10%−70% of the total microbeam length of 140 μm, and 25%−50% of the 25 μm microbeam thickness, but a uniform height of 50 μm in the z-direction. These machines were created using a UV exposure time ranging from 80 ms to 150 ms; thus, our fabrication process can produce a single batch of up to 500 micromachines in just three minutes (see **Fig. 1e**). Further details on our fabrication process can be found in the Materials and Methods section.

We then characterized these shape-transformable micromachines in an experimental setup consisting of a piezo transducer (PZT) bonded onto a glass slide placed adjacent to an acoustic liquid chamber. The setup was mounted on an inverted microscope, and the transient dynamics of the micromachines were studied using a high-speed camera. We applied acoustic fields with operating frequencies of 5.0−8.0 kHz and voltage amplitudes of 10−60 $V_{PP}$. Further details are provided in the Materials and Methods section.

The soft hinge part of the acoustic-deformable microbeam serves as a crucial component in our transformable micromachines, enabling their shape-shifting capabilities. As sound propagates through the liquid medium and encounters the micromachine, the untethered microscale auxetic contracts in both length and width, achieving its fully folded shape in 30−200 ms (see Supplementary Video 1). A representative example of the micromachine is illustrated in **Fig. 1f**: the left square geometry of the micro auxetic denoted as S1 undergoes a transformation that results in the two smaller squares labelled S1a and S1b. Similarly, as shown in **Fig.1g**, a quadrupedal microrobot demonstrates selective folding of its outer four square-shaped legs, which leads to shape transformation and division of Leg 1 into smaller squares (Leg1a and Leg1b), completed in 50 ms (see Supplementary Video **2**). This dynamic and selective folding is controlled by acoustic voltage, which can be modulated by an external function generator, allowing the microstructures to exhibit a walking gait. Upon switching the acoustic field off, both the micromachine and microrobot swiftly revert to their original shapes (see Supplementary Fig. S2**)**.

To gain further insights into the mechanism underlying the shape transformation (**Fig. 1h)**, we observed the dynamics of acoustic-deformable microbeams in an acoustic field through high-speed microscopy. As shown in **Fig. 1i**, our observations unveil a fascinating phenomenon: when the microbeam is clamped-clamped constrained and subject to an acoustic field, the soft hinge oscillates periodically (**Fig. 1j**). We discovered that the sound field specifically concentrates on the soft hinge region, resulting in amplified amplitude oscillation of the hinge compared to the other parts; in fact, its oscillation is ~5 times larger than that of the rigid link (**Fig. 1k** and Supplementary Video 3). Injection of tracer microparticles into the liquid revealed the fast oscillations of the soft hinge part to create vortices in the surrounding liquid, commonly known as microstreaming. Additionally, the oscillation generates a localized acoustic radiation force from the soft hinge. Collectively, the concentration of forces exerted at the soft hinge, which arises from the combined effects of acoustic radiation and microstreaming, is predominantly counterbalanced by the stored elastic strain energy within the microbeam body during each oscillation cycle. When ends of the microbeam are not clamped but have damping elements arising from the viscous fluid, friction, and interconnect effect from other beams, termed the "damped-damped" condition, we observed the oscillation of the soft hinge to induce rotational motion of the rigid link, which in turn initiated folding of the microstructure (**Fig. 1i** and Supplementary Video 3). The angle between the two rigid links decreases linearly with increasing acoustic excitation time, as depicted in **Fig. 1j**. This behaviour is attributable to the torque resulting primarily from the difference in stiffnesses between the rigid link and soft hinge.

We performed a series of experiments to optimize a single unit of an acoustic-deformable microbeam, varying parameters such as the dimensions of the soft hinge (thickness and length), the UV exposure time, and the acoustic excitation voltage with the goal of facilitating efficient transformation. We found that reducing the thickness of the soft hinge by 50% (from 25 μm to 12.5 μm) led to a 60% increase in the oscillation amplitude ($\varepsilon$) of the clamped-clamped microbeam at the maximum voltage tested. Similarly, increasing the length of the soft hinge by 60% resulted in a substantial 40% amplification of the oscillation amplitude. In addition, we observed that micromachines fabricated with longer UV exposure time exhibited lesser oscillation amplitude (refer to Supplementary Fig. S3 for characterization graphs).

**Figure 1l** shows the Young's modulus of the acoustic-deformable microbeam, measured using an atomic force microscope (AFM) (see Supplementary Note 1 and Supplementary Fig. S4 for details). The results indicate that increasing the UV exposure time results in an overall increase in the stiffnesses of the entire microbeam, with particularly steep increases after 150 ms. Moreover, at high exposures, the stiffness differential between the rigid link and the soft hinge is substantially greater. The wide range of options for modulating stiffnesses, 50–270 MPa for the link and 10–50 MPa for the hinge, provides exciting new possibilities for designing and programming transformable micromachines and microrobots.

The variable-stiffness design enables acoustic-deformable microbeams to adopt a linearized model to quantify the oscillation and folding dynamic behaviour in response to acoustic excitation,

$$\boldsymbol{Q}(x,t) = \boldsymbol{F}(V,t,x) \cdot \sum_{i=1}^{3} \boldsymbol{N}(x)\boldsymbol{q}^i(t), \quad (i=1,2,3\ldots) \qquad (1)$$

where $\boldsymbol{Q}(x,t)$ is a time-varying deflection of the entire microbeam at position $x$ along its long axis. The acoustic forces on this microbeam, denoted $\boldsymbol{F}(V,t,x)$ with the acoustic excitation voltage $V$, predominantly comprise the acoustic radiation force $|F_r|$ and the acoustic streaming force $|F_s|$. The acoustic radiation force is calculated as $|F_r^i| = \int_{\Omega^i} \frac{2\alpha I^i(x,y)}{c}$, showing a relationship with the absorption coefficient $\alpha$, the temporal average intensity $I$ of the acoustic wave at the given location, and the speed of the sound in the microbeam $c$, where $i$ denotes the $i$th composing element of the microbeam with different thicknesses, stiffnesses, and lengths. Furthermore, the acoustic streaming force is represented as $\left|F_s^i\right| \approx \int_{\Omega_s^i} \tilde{\alpha} \boldsymbol{v}_0^2 \, exp(-2\tilde{\alpha} r)$, where $\Omega_s$ is the integral domain of the microbeam's surface, $\tilde{\alpha}$ is the acoustic attenuation coefficient, $\boldsymbol{v}_0$ is the characteristic source velocity amplitude, and $r$ is the propagation distance of the sound wave. $\boldsymbol{N}(x)$ is the mode function matrices of the microbeam which is governed by the beam's boundary conditions. The generalized coordinate vector $\boldsymbol{q}^i(t)$ serves as a comprehensive representation that captures the entire time-varying displacement. The model aids in quantifying the displacement of the microbeam and supports a theoretical understanding of its dynamic behaviour, influenced by factors including its boundary conditions, acoustic excitation, and structural design. This model further suggests that reducing the hinge thickness, increasing the hinge length, and identifying appropriate UV exposure times are the key factors in maximising the oscillation amplitude and achieving efficient transformation, which is consistent with the results of microbeam characterisation experiments. The model can be adapted to more microbeam configurations and acoustic excitation conditions, such as multiple soft hinges, different boundary conditions, and variable acoustic propagation directions, which will facilitate the design of new acoustic-transformable micromachines. Detailed theoretical development, verification, and simulations of the model are provided in Supplementary Note 2 and Fig. S5.

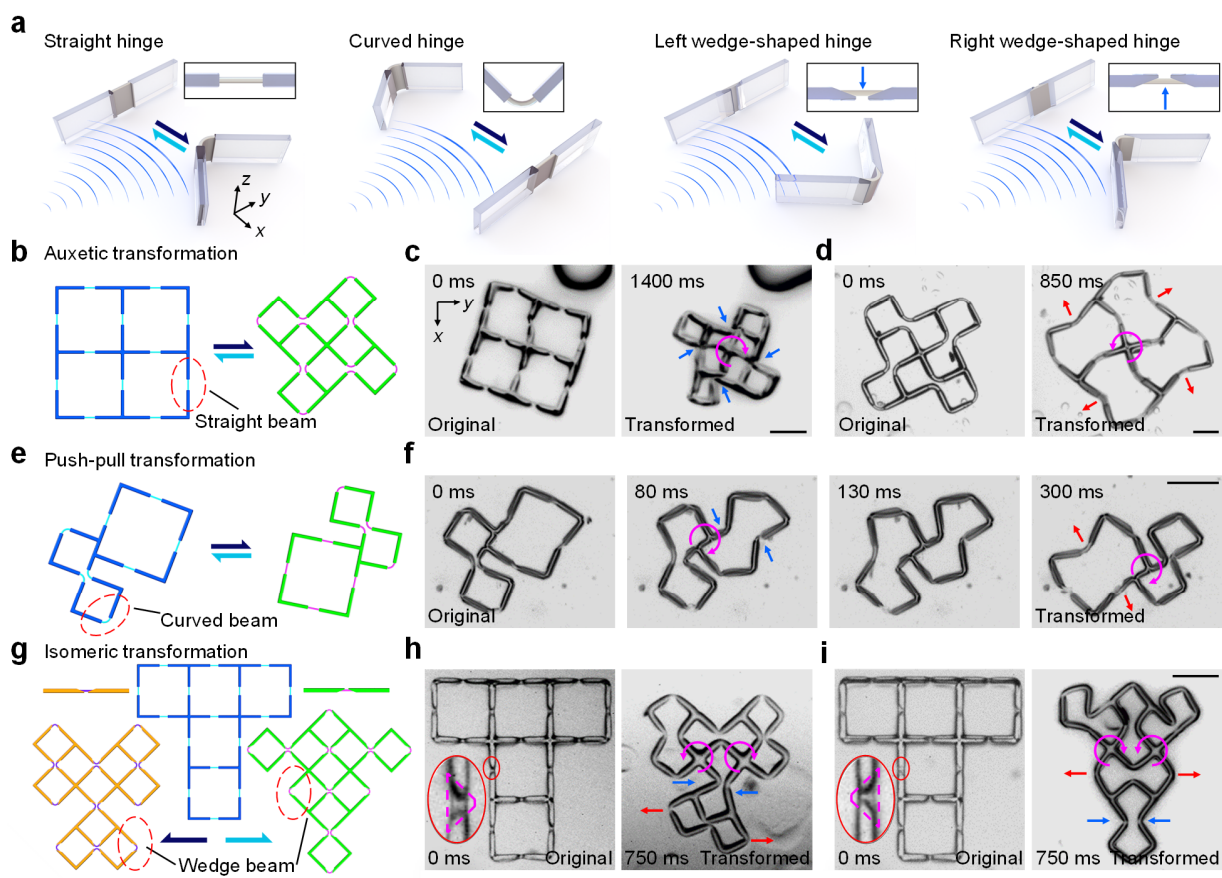

**Fig. 2| Versatile shape transformation demonstrated by various soft micromachines. a**, Four types of acoustic-deformable microbeams and their deformed shapes with an acoustic activation. Insets show the top view of microbeams. The blue arrows demonstrate the predetermined folding directions of the wedge-shaped soft hinges, respectively. **b**, Bilateral auxetic transformation. **c**, A four-square micromachine constructed from straight acoustic-deformable microbeams folds into its full-folded shape. **d**, An inverse four-square micromachine constructed from curved acoustic-deformable microbeams is extended to its full-extended shape. Blue and red arrows respectively denote the folding and unfolding behaviour of the microbeams. Magenta curved arrows show the rotation direction of the centre cross of the micromachine. **e**, **f**, Push-pull transformation showcased by a micromachine constructed from both straight and curved acoustic-deformable microbeams. Upon acoustic activation, the folded structure extends when the extended structure folds. **g**, Isomeric transformation showcased by T-shaped micromachines. **h**, a T-shaped micromachine with the left wedge microbeam transforms toward a "concave" shape. **i**, a T-shaped micromachine with the right wedge microbeam transforms toward a "convex" shape. Scale bar (for all optical images), 100 μm.

**Versatile Shape Transformation of Micromachines**

We designed various types of acoustic-transformable micromachines by incorporating acoustic-deformable microbeams with a range of straight, curved, and wedge-shaped soft hinges, as shown in **Fig. 2a**. The curved microbeams demonstrate similar deformation behaviour to straight microbeams; however, in curved microbeams, the oscillating hinge system unfolds and extends. We have further developed specialized asymmetric wedge-shaped hinge can fold in the predetermined direction (see Supplementary Fig. S6). Micromachines incorporating these respective hinges exhibited different behaviours when acoustically actuated. The four-square micromachines, constructed from straight microbeams, contract upon acoustic activation, as illustrated in **Fig. 2, b** and **c**. Conversely, micromachines composed of curved microbeams extend when subjected to acoustic stimulation, shown in **Fig. 2d** and Supplementary Video 4. As these microstructures initiate folding or unfolding, they start to rotate around their centre cross (x). When folding, the auxetic structure rotates clockwise, and when unfolding, it rotates counterclockwise. The torque driving these rotations can be attributed to the asymmetric positioning of the soft hinges, which produces an unbalanced force when subject to acoustic streaming and radiation forces (Supplementary Fig. S7). In addition, as the deformable microbeams are interconnected, any torque applied to one will affect the others. Thus, the resultant net force and net torque cause the micromachines to undergo transformation with a negative Poisson's ratio of around −1 (refer to Supplementary Note 3, Fig. S8, and Fig. S9 for calculation method and results). These micromachines possess a shear modulus significantly larger than their elastic modulus. Consequently, when subjected to an acoustic field, they exhibit consistent and stable auxetic transformation, while maintaining resistance to shear deformation. (see Supplementary Note 4 and Table 1 for optimization details of these auxetic micromachines).

**Figures 2, e** and **f** demonstrate the push-pull transformation of a micromachine incorporating both straight and curved acoustic-deformable microbeams. When we activate the acoustic field, different parts of the micromachine extend and contract simultaneously (see Supplementary Video 4). The combined use of different microbeams allows diverse designs of acoustic-transformable micromachines. We have also developed a T-shaped micromachine that can undergo isomeric transformation. This capability is achieved through the symmetric placement of soft hinges: each hinge produces an unbalanced force and torque, and this simultaneous force coming from two directions causes isomeric transformation (**Fig. 2g**). To control the direction of folding, we incorporated the asymmetric wedge-shaped soft hinge into the design. This specialized hinge generates a net force in a predetermined direction; specifically, a T-shaped micromachine with a left wedge-shaped hinge deforms outward into a "concave" shape **(Fig. 2h** and Supplementary Video 4) while one with a right wedge-shaped hinge folds inwards into a "convex" shape **(Fig. 2i** and Supplementary Video 4). Through this strategic design of hinge configuration, we have effectively engineered the micromachine to exhibit predictable and targeted transformation behaviour, supported by our numerical simulations (see Supplementary Fig. S10).

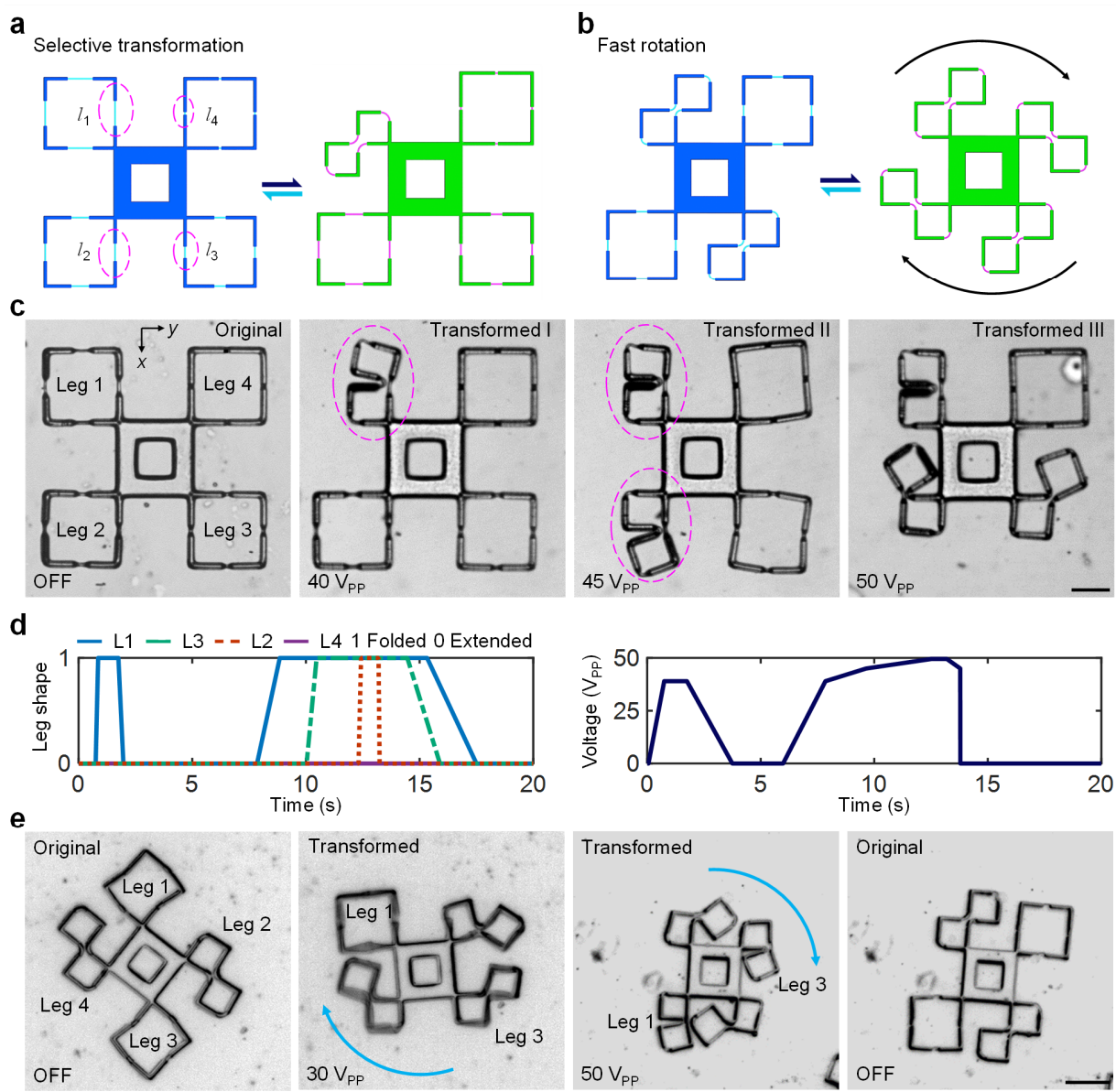

**Fig. 3| Selective shape transformation tuned by acoustic excitation voltages. a**, Selective transformation of a quadrupedal microrobot with square-shaped legs constructed from straight microbeams with different hinge lengths, namely, 46.2, 33.0, 19.8 and 6.60 μm. **b**, Fast rotation of a microrobot constructed from straight and curved microbeams with the diagonally positioned extended legs and folded legs. **c**, The legs of a microrobot are selectively folded as the acoustic excitation voltage is gradually increased to 40, 45, and 50 $V_{PP}$. The magenta dotted ellipse shows the folded leg. **d**, Plot showing the dynamic folding of the microrobot's legs when tuning the acoustic excitation voltage against time. **e**, A microrobot exhibits clockwise rotation in an acoustic field. As the excitation voltage increases, the microrobot shows a higher degree of contraction and exhibits a higher rotational speed. The blue curved arrow shows the rotational direction. Scale bar (for all optical images), 100 μm.

**Selective Shape Transformation of Microrobots**

To show the selective shape transformation capacity of micromachines by adjusting the acoustic excitation voltage, we have developed quadrupedal microrobots modelled after a tetrapod consisting of a central bulky body and four surrounding legs, as shown in **Fig. 3, a** and **b**. These legs are constructed using various incorporations of straight and curved acoustic-deformable microbeams. **Fig. 3a** shows a microrobot with four square-shaped legs designed with straight microbeams of different hinge lengths. The microrobot's shape is selectively transformable, as shown in **Fig. 3c**. For excitation voltages below 30 $V_{PP}$, the microrobot retains its original shape. At 40 $V_{PP}$, the leg with the longest hinge undergoes folding. When the voltage is further increased to 45 $V_{PP}$, leg 2 folds while leg 1 remains folded. Finally, when the voltage reaches approximately 50 $V_{PP}$, leg 3 folds with folded legs 1 and 2. Leg 4, which contains the shortest hinge, does not exhibit folding within the current acoustic excitation (see Supplementary Video 2). **Fig. 3d** and Supplementary Fig. S11 shows the dynamic and flexible shape transformation process when adjusting the acoustic excitation voltage. The folding sequence of the microrobot's legs corresponds to the characterization and model analysis of the straight microbeams. As the length of the soft hinges increases, the oscillation becomes easier to induce, thereby requiring lower voltages to initiate folding. This tunability enables us to achieve addressable and programmable shape transformation, highlighting its potential for developing sophisticated and multifunctional microrobots.

We also designed a microrobot consisting of a pair of square-shaped legs with straight hinges and a pair of folded-square-shaped legs with curved hinges arranged diagonally (**Fig. 3b**). Remarkably, when subjected to the acoustic stimulation, the microrobot displays rotational motion (**Fig. 3e** and see Supplementary Video **5**), which can be attributed to the net torque generated by its asymmetric structure. We observed a significant increase in the rotational speed with higher acoustic excitation voltages, which increased around13 times when increasing the voltage from 20 to 50 $V_{PP}$. For voltages below 20 $V_{PP}$, the microrobot exhibits a low rotational speed, ranging from 0 to 30 revolutions per minute (rpm), while retaining their original shape. As the voltage increases to 30 $V_{PP}$, the four legs exhibit the dynamic shape changes between folding and unfolding during rotation. We assume this is because the applied acoustic forces are similar to elastic strain energy, resulting into a dynamic balance. When the excitation voltage exceeds 45 $V_{PP}$, the acoustic forces are larger than the elastic strain energy and result into the full-folded shape of all the four legs, causing a lower inertia moment and a faster rotational speed, with a speed around 400 rpm for 52.5 $V_{PP}$ (see Supplementary Fig. S12 for the rotational speed graph). By using these acoustic-deformable microbeams as the foundational structural module, a broad range of microrobots can be designed with various choices in structural configurations, geometric dimensions, and stiffnesses (see Supplementary Fig. S13 for more robotic designs).

**Discussion**

The interaction between sound waves and soft matter, in particular materials of different stiffnesses, remains largely unexplored. In our study, we present the first demonstration of acoustically activated ultrafast transformable soft micromachines. These micromachines incorporate deformable soft microbeams that allow them to change shape in milliseconds upon the application of an acoustic field, as well as to revert to their original state equally rapidly when the field is deactivated. Our simulation and theoretical studies further confirmed this fundamental new shape-transforming mechanism. It merits mention that these micromachines exhibit remarkable reliability and adaptability. Adopting extreme geometric designs such as curved links, point hinges, block links and broken portions, these soft micromachines still achieve ultrafast shape transformation (Supplementary Fig. S14 and see Supplementary Video 6). As a new mechanism of shape transformation, acoustic actuation enables miniaturisation at micro and nanoscales, with rapid prototyping, compact size, remote and wireless operation, and ultrafast and reversible responses.

The development of versatile deformable microbeams and shape-transformable micromachines paves the way for exciting advancements in multiple scientific domains. From a physics perspective, they offer an exciting opportunity to explore acoustic interactions with soft matter, allowing for the exploration of, for example, oscillation modes and synchronisation effects (Supplementary Fig. S15). In the field of microdevices and microfluidics, integrating our micromachines into chip platforms facilitates the use of such platforms in drug delivery, object loading and various mechanical tests of living or non-living matter, such as fatigue testing (Supplementary Fig. S16). In the realm of metamaterials, designs incorporating acoustic-deformable microbeams can enable intricate 3D shape morphing and the creation of 2D-to-3D origami/kirigami microstructures. These structures can be used for optical guidance, acoustic guidance and energy amplification, among other purposes. In robotics, acoustically deformable beams open up novel micro-scale design paradigms and can enhance manoeuvrability by enabling functionalities such as steering, propulsion and environmental adaptation. Furthermore, in flexible electronics, shape-transformable micromachines are beneficial for safe, portable and friendly interactions, as well as for developing new functions. For medicine, as acoustic fields can easily extend into the body, these micromachines have the potential to dynamically navigate through complex blood vessels of varying sizes in the immediate future.

Our study shows exciting prospects across multiple scientific domains, but limitations currently exist when multiple micromachines are simultaneously actuated in an acoustic field. In such scenarios, micromachines tend to attract each other, leading to potential changes in their movement directions. This phenomenon is primarily caused by the strong secondary force of acoustic radiation resulting from the interaction of scattered waves between adjacent micromachines. However, the understanding of this complex acoustofluidic theory is incomplete. Different materials with varying acoustic impedances can be used to fabricate micromachines to fine-tune the secondary acoustic radiation force, enabling control over the attraction between micromachines. This approach also opens up possibilities for developing novel self-assembly mechanisms.

## Methods

### Photosensitive mixture

The utilized photosensitive mixture is composed of a synthetic polymer (Polyethylene glycol with a molecular weight of 700, PEG 700, Sigma-Aldrich) and a photo-initiator (2-hydroxy-2-methyl-1-phenyl-propan-1-one, Darocur 1173, Sigma-Aldrich) at a ratio of 5:1. The photo-initiator will trigger the crosslink of PEG 700 when exposed to UV light. One droplet (~50 µl) of Rhodamine B solution (Sigma-Aldrich) is added to the mixture to align the mask, and one droplet (~50 µl) of blue food dye is added to the solution to visualize the fabricated micromachines.

### Fabrication method

Our custom-built UV photopolymerization is developed on an inverted microscope (NIKON, Eclipse Ti). A UV lamp (Nikon Intensilight C-HFGI) and a shutter controller (Vincent Associates, VCM-D1) are connected to irradiate a high-resolution photomask (CAD/Art Services, Inc.) inserted into the field stop of the microscope. 20−30 µL of the photosensitive mixture is spread on the fabrication area of a glass substrate (24×60×0.15 mm), and then the mixture is flatted by a cover glass (24 × 24 × 0.15 mm). When the mask is irradiated, passed UV light is focused by a 20x objective to polymerize the photosensitive mixture. The exposure time varied in a range of 50−300 ms. Notably, if the UV exposure time is too short, the structures may fail to polymerize or be excessively soft, and consequently break when exposed to high excitation voltages. Hereby one micromachine is produced according to the photomask pattern. Once a micromachine is fabricated, we move the platform of the microscope for the next fabrication. Finally, the microstructures undergo a cleaning process to remove any residual contaminants by using isopropanol (IPA).

### Acoustic setup

The experimental setup is built on a thin glass substrate, and a circular piezoelectric transducer (27×0.54 mm, resonance frequency 4.6 kHz ± 4%, Murata 7BB-27-4L0) is affixed to the glass substrate using an epoxy resin (2-K-Epoxidkleber, UHU Schnellfest). A droplet containing ~50 µL of deionized water is spread on the substrate and covered with a cover glass to serve as a thin liquid medium. The substrate is then mounted on an inverted microscope (Axiovert 200M, ZEISS). An electronic function generator (AFG-3011C, Tektronix) and an amplifier (0−60 $V_{PP}$, 15x amplification, High Wave 3.2, Digitum-Elektronik) are connected to the transducer to generate sound waves with tuneable excitation frequencies and voltages.

## Imaging and analysis

The shape transformation of micromachines is recorded with a high-speed camera (Chronos 1.4, Kron Technologies) attached to the inverted microscope. Recording frame rates range from 50 to 32,668 frames per second (fps). Recorded footage is analysed in ImageJ.

## Acknowledgments

We thank Vincent Christopher Winderoll, Jannes Huber, and Sam Rahimi for the assistance of experiments. This project has received funding from the European Research Council (ERC) under the European Union's Horizon 2020 research and innovation programme grant agreement No. 853309 (SONOBOTS). Z.Z. acknowledges financial support from the China Scholarship Council (202006210065).

## Author contributions

D.A. conceived and supervised the project. Z.Z. developed the design and fabrication of micromachines, performed experiments and data analysis. Z.S., Z.Z., and D.A. contributed to the theoretical understanding. Z.S., and Z.Z. contributed to the numerical simulations. Z.Z., Z.S., and D.A. contributed to the scientific presentation, discussion, and manuscript revision.

## Competing interests

Authors declare that they have no competing interests.

## Data and materials availability

All data are available in the main text or the supplementary materials.

## Supplementary materials

Supplementary Notes 1— 4
Supplementary Figs. S1 — S16
Supplementary Table 1
Supplementary References 1— 7
Supplementary Videos 1— 6

# Supplementary Materials for

## SonoTransformers: Ultrafast Transformable Wireless Microscale Machines

Zhiyuan Zhang, Zhan Shi, and Daniel Ahmed*

Acoustic Robotics Systems Lab, Institute or Robotics and Intelligent Systems, Department of Mechanical and Process Engineering, ETH Zurich; Rüschlikon, CH-8803, Switzerland

*Corresponding author: dahmed@ethz.ch

**The PDF file includes:**

Supplementary Notes 1─ 4
Supplementary Figs. S1 ─ S16
Supplementary Table 1
Legends for Supplementary Videos 1 ─ 6
Supplementary References 1─ 7

**Other Supplementary Materials for this manuscript include the following:**

Supplementary Videos 1─ 6

## Supplementary Notes

**Note 1. Estimation of micromachines' Young's modulus.** We utilized the Atomic Force Microscope (AFM) to test the Young's modulus. In the test mode of Peak Force Mapping in Air, we set the scanning size of the sample at 500 nm and the scanning rate at 1.95 Hz. The cantilever parameters include the spring constant of 42 N/m, the tip radius of 8 nm, and the tip opening angle of 25°. During scanning, the relationship of the force against *z*-axial displacement of the tip was monitored and captured (as shown in **fig. S5**). Then, based on the slop shown by the force-displacement graph and Sneddon model [1, 2], we calculated the Young's modulus $E_s$ of micromachines as

$$E_s = \frac{F\pi(1-v_s^2)}{2\tan\alpha_t \cdot \delta^2} \tag{S1}$$

where $v_s$ is Poisson's ratio of the sample, which was estimated as 0.33. $F$ is the measured force, $\alpha_t$ is the opening angle of the tip, and $\delta$ is the deformation distance of the tip. We tested the Young's modulus of these micromachines with different exposure times ranging from 50 ms to 300 ms with an incremental step of 50 ms, respectively.

**Note 2. Modeling of the acoustic-deformable microbeam.** We developed a theoretical model that can describe the deformation of each mass point on the straight acoustic-deformable microbeam under acoustic activation, as shown in **Note Extended Fig. N1**. The microbeam is first modeled separately with each comprised element based on the boundary condition and subsequently assembled based on the displacement and stress continuity at the connection interface between the soft hinge and rigid link.

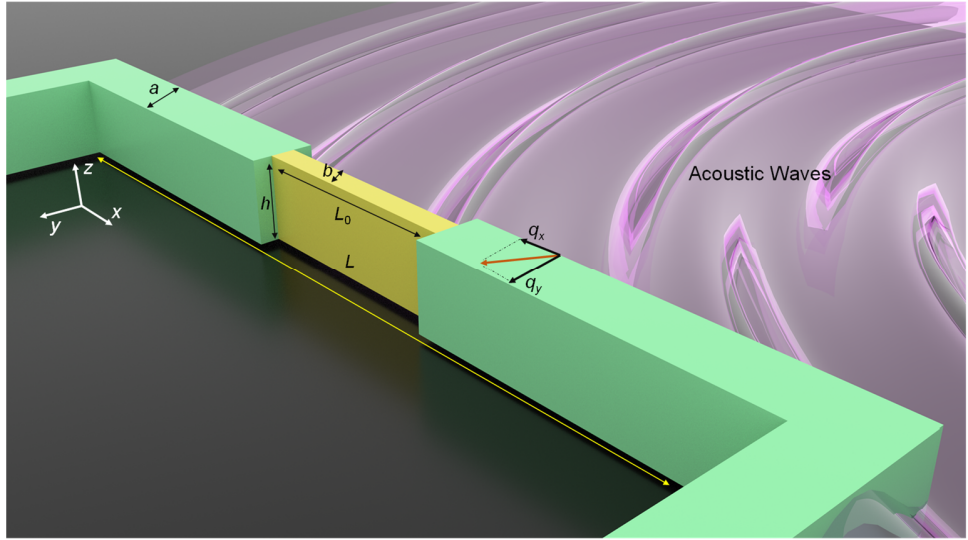

**Fig. N1| Illustration of the interaction between acoustic waves and the acoustic-deformable microbeam.** The acoustic waves can propagate from all directions. The interconnected microbeam has the "damped-damped" boundary condition. The soft hinge has different stiffnesses and geometric parameters compared with the rigid link.

For the strain energy modeling, as shown in **Note Extended Fig. N2**, $P$ is one of the mass points on the centroid line of the microbeam, and $p$ is one mass point on the cross-section which is not on the centroid line. The undeformed position vector $\boldsymbol{R_0}$ and the deformed position vector $\boldsymbol{R_d}$ of the mass point $p$ are

$$\boldsymbol{R_0} = \begin{bmatrix} x \\ y \end{bmatrix} \tag{S2}$$

$$\boldsymbol{R_d} = \begin{bmatrix} x + u - \sin(\theta)\, y \\ v + \cos(\theta)\, y \end{bmatrix} \tag{S3}$$

where $x$ and $y$ are the axial (along the microbeam) and vertical (perpendicular to the microbeam's centroid axis) locations of point $p$. $u$ and $v$ are the axial and transverse displacements of $P$. $\varphi$ is the rotation angle of the centroid axis of the deformed microbeam, $\psi$ is the additional rotational angle because of the shear deformation, thus the total rotation angle of the cross-section $\theta$ can be expressed as $\theta = \varphi - \psi$. Since in experiments, the deflection dominates the microbeam deformation while the extension and rotation are small and neglectable, the axial and shear strains of the microbeam are assumed to be small. Thus, Green's strain tensor [3] still applies to the strain vector of the microbeam which meets the following equation,

$$d\boldsymbol{R}_d^T d\boldsymbol{R}_d - d\boldsymbol{R}_0^T d\boldsymbol{R}_0 = 2[dx \quad dy]\begin{bmatrix} \varepsilon_{xx} & \varepsilon_{xy} \\ \varepsilon_{yx} & \varepsilon_{yy} \end{bmatrix}\begin{bmatrix} dx \\ dy \end{bmatrix} \quad (S4)$$

where $\varepsilon_{xx}$, $\varepsilon_{xy}$, $\varepsilon_{yx}$, and $\varepsilon_{yy}$ are the strain tensors. $\varepsilon_{yy} = 0$ is satisfied for the slender shape beam according to the uniaxial stress assumption [4]. The strain energy $U$ of the $i$th element can be expressed as follows

$$U^i = \iiint_{\Omega^i} \frac{1}{2}\boldsymbol{\varepsilon}^T \boldsymbol{E} \boldsymbol{\varepsilon} \, dS dx \quad (S5)$$

where $\Omega^i$ denotes the integral domain of the $i$th element. $S$ is the area of the microbeam's cross-section. The strain tensor matric $\boldsymbol{\varepsilon}$ can be obtained from Eq. (S4) and the modulus parameters matric $\boldsymbol{E}$ is given as

$$\boldsymbol{\varepsilon} = \begin{bmatrix} \varepsilon_{xx} \\ \gamma_{xy} \end{bmatrix} = \begin{bmatrix} 1 & y & 0 & y^2 \\ 0 & 0 & 1 & 0 \end{bmatrix}\begin{bmatrix} u' + \frac{1}{2}(v'^2 + u'^2) \\ (1+u')\theta' \cos(\theta) - v'\theta' \sin(\theta) \\ v'\cos(\theta) - (1+u')\sin(\theta) \\ \theta'^2 \end{bmatrix} \quad (S6)$$

$$\boldsymbol{E} = \begin{bmatrix} E(x,y) & 0 \\ 0 & G(x,y) \end{bmatrix} \quad (S7)$$

where $\gamma_{xy}$ is the planar shear strain. The prime " $'$ " denotes the derivative with respect to $x$. $E(x,y)$ and $G(x,y)$ are the Young's modulus and shear modulus. The variation in strain energy of the $i$th element is expressed as

$$\delta U^i = \iiint_{\Omega^i} \delta \boldsymbol{\varepsilon}^T \boldsymbol{E} \varepsilon dS dx \tag{S8}$$

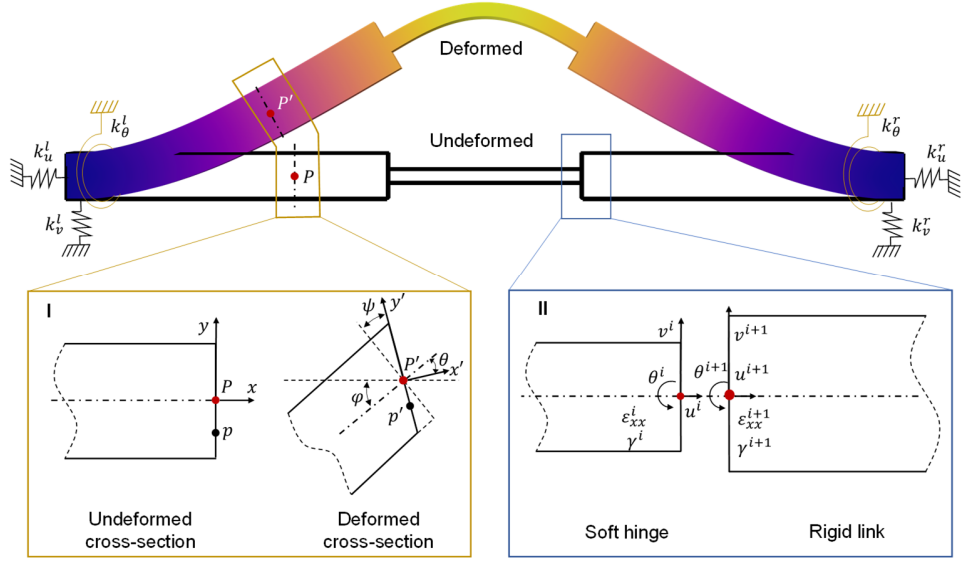

**Fig. N2| Modeling of the straight acoustic-deformable microbeam.** Inset I shows the letter indication of the mass point $p$ before and after deformation. Inset II shows the continuity condition of the connection interface between the rigid link and soft hinge.

Under the acoustic activation, the microbeam undergoes time-varying oscillation thus, the kinetic energy of the $i$th element can be expressed as

$$T^i = \iiint_{\Omega^i} \frac{1}{2} \dot{\boldsymbol{R}}_d^T \rho \dot{\boldsymbol{R}}_d dS dx \tag{S9}$$

in which $\rho$ is the density of the microbeam. '$\dot{\boldsymbol{R}}$' denotes the first derivative of the position vector versus time. The variation in kinetic energy can be written as

$$\delta T^i = -\iiint_{\Omega^i} \delta \boldsymbol{R}_d^T \rho \ddot{\boldsymbol{R}}_d dS dx \tag{S10}$$

The acoustic forces predominantly origin from the following two parts: the acoustic radiation force $F_r$ and the acoustic streaming force $F_s$. The acoustic radiation force on the $i$th element is obtained as

$$\left|F_r^i\right| = \int_{\Omega^i} \frac{2\alpha I^i(x,y)}{c} \tag{S11}$$

where $\alpha$ is the absorption coefficient, $I(x,y)$ is the temporal average intensity of the acoustic wave in the given location, $c$ is the sound speed in water [5]. The acoustic streaming force radiated from the $i$th element's oscillation surface is estimated through surface velocity integral as

$$\left|F_s^i\right| \approx \int_{\Omega_s^i} \tilde{\alpha} v_0^2 \exp(-2\tilde{\alpha} r) \tag{S12}$$

where $\Omega_s$ is the integral domain of the microbeam's surface, $\tilde{\alpha}$ is the acoustic attenuation coefficient, $v_0$ is the characteristic source velocity amplitude, and $r$ is the distance between the surface of the microbeam and the acoustic source along the sound direction[5]. Considering all acoustic field effects, such as wall reflections and dissipation due to impedance, would be prohibitively expensive. However, we can simplify the modeling by describing all acoustic forces as the combined effects of extension, transverse, and rotational forces. For the complicated acoustic fields, we adopt the following assumptions: The acoustic radiation force applied on the whole microbeam can be regarded as a periodic uniform load, and its effects are decomposed into three forces, the force to extend the beam $f_u$, the force to deflect the beam $f_v$, and the force to rotate the beam $f_\theta$. According to experiments, soft hinges with a relatively small thickness oscillate with relatively large amplitudes (5−50 μm) compared with rigid links (0.1−10 μm), which generate stronger acoustic streaming. In addition, the secondary radiation force between two soft hinges may also play a role when the two hinges deform close to each other. Thus, we adopt the assumptions that the streaming forces and the secondary radiation forces are approximated as the concentrated forces applied on the mid-point of the soft hinge. Consequently, the work done by external acoustics can be expressed as the product of generalized displacement quantities and their corresponding force components.

$$W^i = \boldsymbol{Q}_k^T \boldsymbol{F}|_{\Omega^i} \tag{S13}$$

where

$$\boldsymbol{Q}_k = [u \quad v \quad \theta]^T$$

$$\boldsymbol{F} = [F_u \quad F_v \quad F_\theta]^T$$

$F_u$, $F_v$, and $F_\theta$ are the respective total decomposed forces according to the effect of extension, deflection, and rotation. Therefore, the variation of the acoustic forces' work is derived as

$$\delta W^i = \delta \boldsymbol{Q}_k^T \boldsymbol{F}|_{\Omega^i} \tag{S14}$$

Regarding the different thicknesses of the microbeam's soft hinges and rigid links, it's necessary to consider the continuity conditions between different elements. There is no offset of the centroid line between the two elements (see **Note Extended Fig. N2 II**), thus the displacement and strain between the two elements meet the continuity as

$$\boldsymbol{R}_r^i(y) = \boldsymbol{R}_l^{i+1}(y) \tag{S15}$$

$$\boldsymbol{E}\boldsymbol{\varepsilon}_r^i(y) = \begin{bmatrix} \varepsilon_{xx}^i \\ \gamma_{xy}^i \end{bmatrix} = \begin{bmatrix} \varepsilon_{xx}^{i+1} \\ \gamma_{xy}^{i+1} \end{bmatrix} = \boldsymbol{E}\boldsymbol{\varepsilon}_l^{i+1}(y) \tag{S16}$$

where $\boldsymbol{R}_r^i$ and $\boldsymbol{R}_l^{i+1}$ denote the displacement vector of the right and left interface, respectively; $l$ and $r$ represent the left and right boundary. $\varepsilon_r^i$ and $\varepsilon_l^{i+1}$ are the corresponding strain vectors for both connection interfaces. Note that the nonlinear relation between the slope angle $\varphi$ of element centroid line and displacement quantities ($u$, $v$) is absent during the derivation of strain energy, kinetic energy, and external acoustics forces work; hence, a constrained variational method is developed to introduce the nonlinear constraints into the energy functional according to the constrained functional method

$$\Pi_n^i = \sum_{i=1}^{3} \int_{t_1}^{t_2} \int_{l_i}^{l_{i+1}} \left( \left( sin(\varphi) - \frac{v'}{\sqrt{\left(1+u'\right)^2 + v'^2}} \right)^2 + \left( cos(\varphi) - \frac{1+u'}{\sqrt{\left(1+u'\right)^2 + v'^2}} \right)^2 \right) dxdt \tag{S17}$$

According to the structural design, the boundary conditions of the interconnected microbeam at both ends are equivalent to three different stiffness, the extension stiffness $k_u$, the transverse stiffness $k_v$, and the rotational stiffness $k_\theta$. The energy functional of the microbeam can be directly written as

$$\Pi_b = \int_{t_1}^{t_2} \left[ \frac{1}{2} (\boldsymbol{Q}_k|_{x=0,y=0})^T \boldsymbol{B}_l + \frac{1}{2} (\boldsymbol{Q}_k|_{x=L,y=0})^T \boldsymbol{B}_r \right] dt,$$
$$\boldsymbol{B}_l = [k_u^l u \; k_v^l v \; k_\theta^l \theta]^T \big|_{x=0,y=0},$$
$$\boldsymbol{B}_r = [k_u^r u \; k_v^r v \; k_\theta^r \theta]^T \big|_{x=L,y=0}.$$
(S18)

Combining the energy functional and the constraint terms, the integrated energy functional is

$$\tilde{\Pi} = \int_{t_1}^{t_2} \sum_{i=1}^{3} (T^i - U^i + W^i) dt - \beta \sum_{i=1}^{3} \Pi_n^i - \Pi_b \tag{S19}$$

where $\beta$ is the constraint weight factor regulating $\Pi_n^i$. According to Hamilton's principle [6], the behavior of the microbeam satisfies the following equation

$$\delta \tilde{\Pi} = \int_{t_1}^{t_2} (\delta T - \delta U + \delta W) dt - \beta \delta \Pi_n - \delta \Pi_b = 0 \tag{S20}$$

To facilitate the calculation of the final governing equations for the microbeam, the assumed mode method is employed to discretize the displacement components [7]. This involves expanding the displacement components into spatial admissible functions and the generalized coordinates

$$\boldsymbol{Q}(x,t) = \sum_{i=1}^{3} \boldsymbol{N}(x) \boldsymbol{q}^i(t), \quad (i = 1,2,3 \dots) \tag{S21}$$

where $\boldsymbol{Q}(x,t)$ is the time-varying deflection amplitude of the microbeam along its axial position $x$. $\boldsymbol{N}(x)$ is the shape function matric along the location $x$ of the microbeam which is determined by the boundary constraints between different elements. The generalized coordinate vectors, $\boldsymbol{q}^i(t)$, serve as a comprehensive representation capable of capturing the entire time-varying displacement.

Finally, the linearized model that quantifies the acoustic-deformable microbeam dynamics is obtained as

$$\boldsymbol{Q}(x,t) = \boldsymbol{F}(V,t,x) \cdot \sum_{i=1}^{3} \boldsymbol{N}(x) \boldsymbol{q}^i(t), \quad (i = 1,2,3 \dots) \tag{S22}$$

where $\boldsymbol{F}(V,t,x)$ denotes the acoustic forces on the microbeam with the acoustic excitation voltage $V$, predominantly comprising the acoustic radiation force $|F_r|$ and the acoustic streaming force $|F_s|$.

We further carried out numerical simulations to verify the microbeam's dynamics. With an isotropic material assumption, parameter matrices, like the Young's modulus and shear modulus, are simplified as constant. The main parameters of the microbeam are assigned the following values:

| Parameter | Value |
| --- | --- |
| Young's modulus | 2e7 Pa |
| Poisson's ratio | 0.3 |
| Shear modulus | 1e4 Pa |
| Total length | 140 μm |
| Soft hinge length | 42 μm |
| $k_u, k_v, k_\theta$ | 1e8 Pa |
| $F_u$ | $1e4\ N/m^2 - 5e4\ N/m^2$ |
| $F_v$ | $1e6\ N/m^2 - 5e6\ N/m^2$ |
| $F_\theta$ | $1e4\ N/m^2 - 5e4\ N/m^2$ |

The theoretical calculations of the centroid line amplitude of the microbeam under different forces, depicted in **Note Extended Fig. N3**, match well with the simulation results performed by COMSOL Multiphysics V6.1, thus verifying the effectiveness of this model. This model allows for a comprehensive characterization of the microbeam's deformation under various boundary constraints and acoustic excitations. It serves as a valuable tool for predicting the microbeam's deformation behavior and offers a methodology for optimizing the structural design of new transformable micromachines.

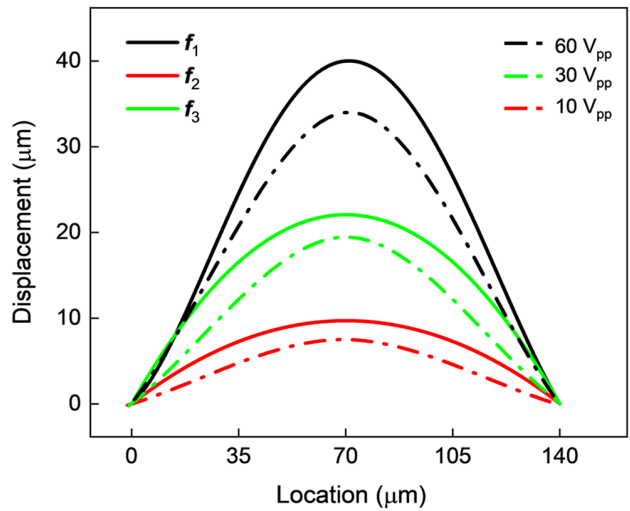

**Fig. N3| Theoretical calculations of the microbeam deformation based on the theoretical model.** The input force matrices $\bm{f} = [F_u\ F_v\ F_\theta]\ (N/m^2)$ are as follows: $\bm{f}_1 = [5e4\ 5e6\ 5e4]$, $\bm{f}_2 = [3e4\ 3e6\ 3e4]$, $\bm{f}_3 = [1e4\ 1e6\ 1e4]$. The simulated deformation configurations of the same microbeam under input voltages of 60 $V_{PP}$, 30 $V_{PP}$, and 10 $V_{PP}$ are shown in the inset, respectively.

**Note 3. Calculation of micromachines' Poisson's ratio.** To calculate the Poisson's ratio, we first measure the dimensions of the micromachines' original shape via the recording footage ($x_0$ and $y_0$ which are obtained along the *x*-axis and *y*-axis, respectively). Then, we measure the dimensions of its transformed shape ($x_1$ and $y_1$). The deformation $\varepsilon_m$ can be obtained by

$$\varepsilon_{mx} = \frac{x_0 - x_1}{x_0}, \varepsilon_{my} = \frac{y_0 - y_1}{y_0} \tag{S22}$$

Based on the deformation along the two dimensions, we then obtain the Poisson's ratio, $v$, as

$$v_m = -\frac{\varepsilon_x}{\varepsilon_y} \tag{S23}$$

**Note 4. Optimization of nine-square acoustic-transformable micromachines.** Several nine-square micromachines were tested with different geometric parameters and excitation voltages to optimize the performance of auxetic shape transformation. When the acoustic excitation voltage is low (<= 10 $V_{PP}$), no shape transformation was observed because the acoustic forces were unable to overcome the elastic strain resistance. As the excitation voltage is larger than 15 $V_{PP}$, some micromachines can deform 10%-70% compared to their full-folded shape because the required reaction force and torque increase with the degree of deformation. With the same exposure time, thinner (<= 10 μm) micromachines will be damaged by high excitation voltage; thicker (>= 100 μm) micromachines cannot deform due to higher stiffnesses. Consequently, as shown in **Table 1**, micromachines with the geometric parameters of hinge lengths ranging from 20 μm to 70 μm and hinge thicknesses ranging from 12.5 μm to 25 μm are capable of achieving full-folded shape transformation when subjected to acoustic excitation frequencies ranging from 5.5 kHz to 6.5 kHz and excitation voltages ranging from 30 $V_{PP}$ to 60 $V_{PP}$. To achieve this transformation behavior, an appropriate UV exposure time of 80 ms to 150 ms is necessary.

**Supplementary Figures**

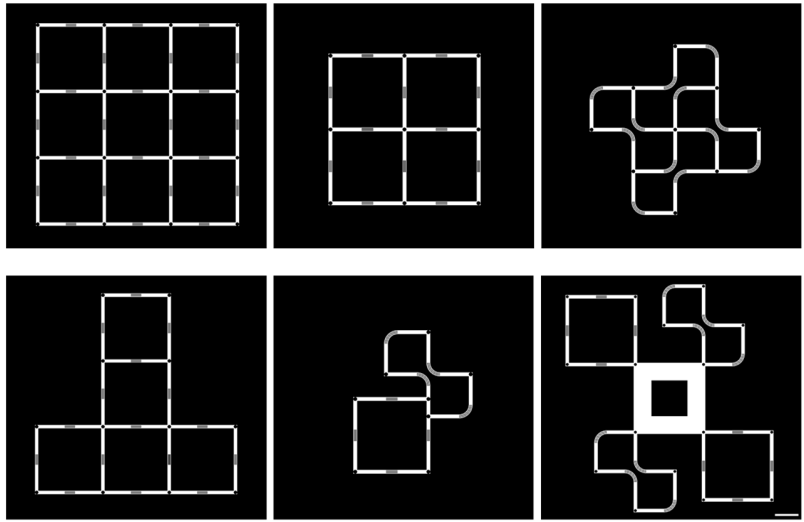

**Fig. S1| Photomasks for various acoustic-transformable micromachines.** The fabricated micromachines are scaled down by a factor of ~16.3 from the dimensions on the masks. Scale bar, 500 μm.

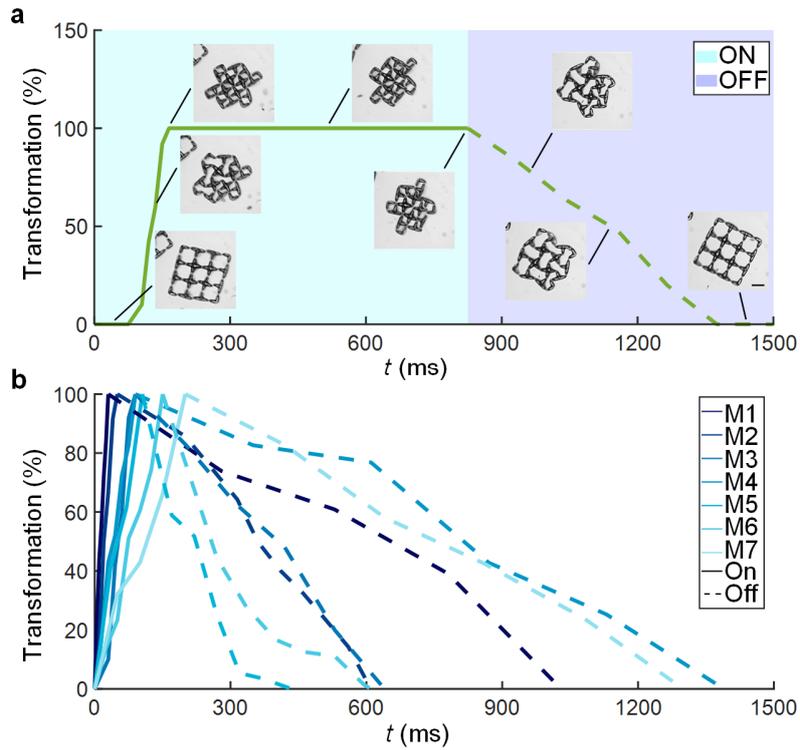

**Fig. S2| Shape transformation of nine-square micromachines against time**. **a**, Plot showing a nine-square micromachine with a UV exposure time of 150 ms contracts toward its full-folded shape in 90 ms. The acoustic excitation voltage and frequency are 45 $V_{PP}$ and 6.5 kHz, respectively. The micromachine holds its folded shape during acoustic excitation. As the acoustic field is deactivated, the micromachine rebounds to its original shape in 550 ms. Scale bar, 100 μm. **b**, Plots showing the shape transformation of seven different nine-square micromachines with UV exposure time 150 ms. As the acoustic field is on, these micromachines contract in 30-200 ms; and they rebound in 400-1000 ms when turning off the acoustic field. The acoustic excitation voltage and frequency are 30 $V_{PP}$-60 $V_{PP}$ and 6.5 kHz, respectively.

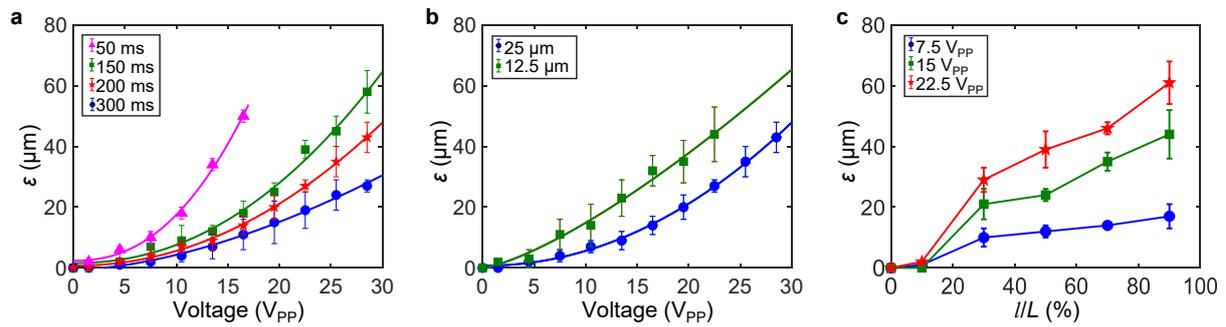

**Fig. S3| Characterization of a single straight acoustic-deformable microbeam. a**, Plots showing the maximum oscillation amplitude of clamped-clamped microbeams with different UV exposure times against the acoustic excitation voltage. **b**, Plots showing the maximum oscillation amplitude of clamped-clamped microbeams with different hinge thicknesses against the acoustic excitation voltage. The UV exposure time is 100 ms. **c**, Plots showing the maximum oscillation amplitude of clamped-clamped microbeams against different hinge lengths to the entire length of the microbeam. The UV exposure time is 100 ms. The acoustic excitation frequency is 6.5 kHz (for all plots).

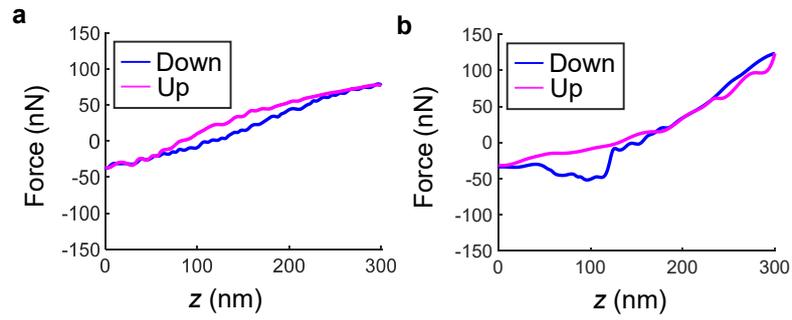

**Fig. S4| Force-displacement graph captured by AFM for the estimation of micromachines' Young's modulus.** Plots of **a**, the soft hinge and **b**, the rigid link with UV exposure time 150 ms.

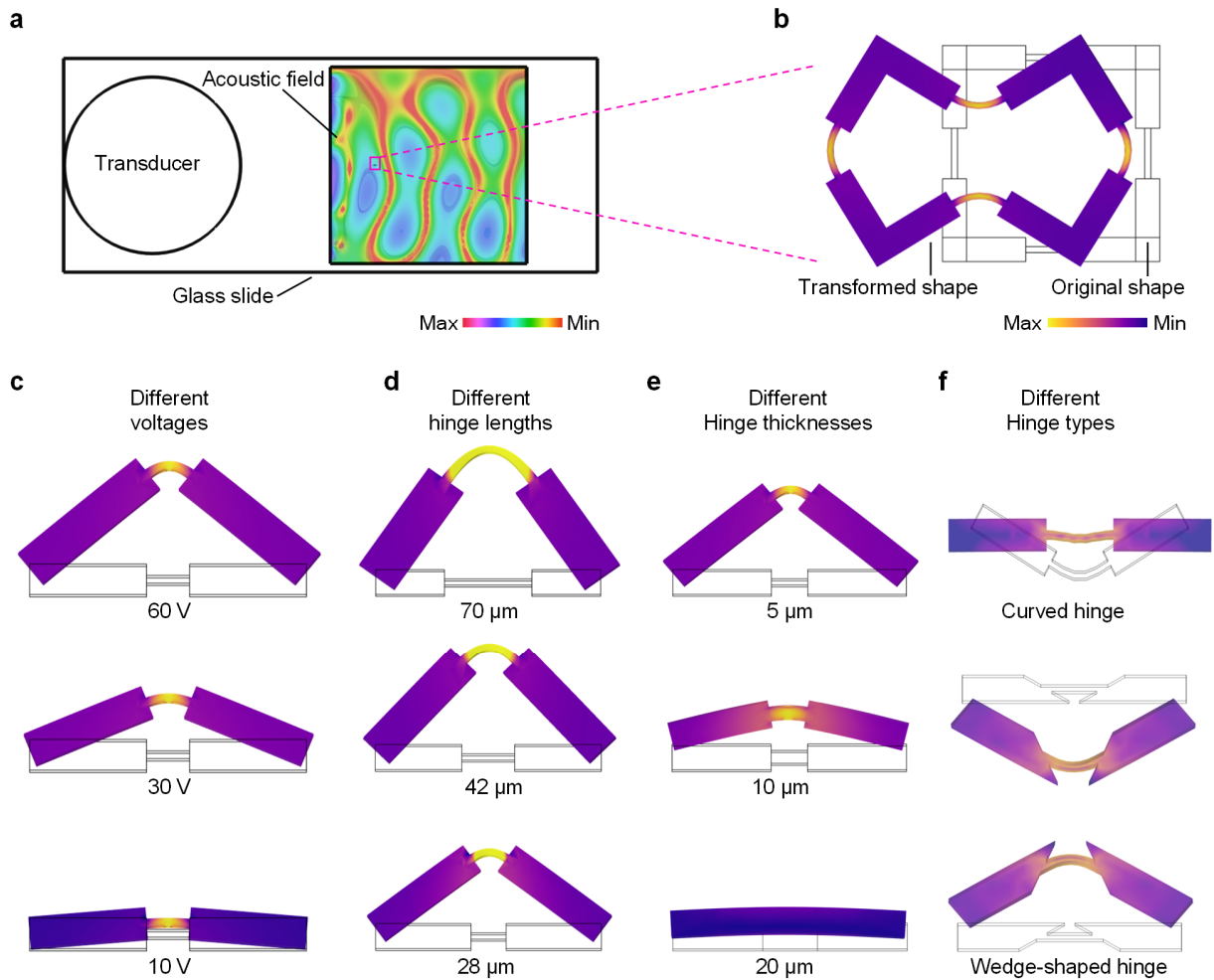

**Fig. S5| Numerical simulation of the acoustic-transformable micromachine and single acoustic-deformable microbeam. a**, Simulation model of the experimental setup with the geometric parameters as the real device. The acoustic pressure under the cover glass is shown. The acoustic frequency and voltage are set to 6.5 kHz and 60 $V_{PP}$, respectively. The micromachine and microbeam are positioned around an area with relatively uniform acoustic pressure in the field. The simulation model was developed with the commercial software, COMSOL Multiphysics V6.1. The color denotes acoustic pressure. **b**, Simulation of a square micromachine with the "free" boundary condition and without friction. The transformed shape is consistent with experimental results (**fig. S8**). The color denotes mechanical stress. **c**, to **f**, Simulations of straight acoustic-deformable microbeams with different acoustic excitation voltages, different hinge lengths, different hinge thicknesses, and different hinge types, respectively. The acoustic voltage is set to 60 $V_{PP}$ for panels **d**, to **f**,. The boundary condition of the microbeam is set to "spring foundation". These simulations correspond with experimental results, confirming the reliability of our findings.

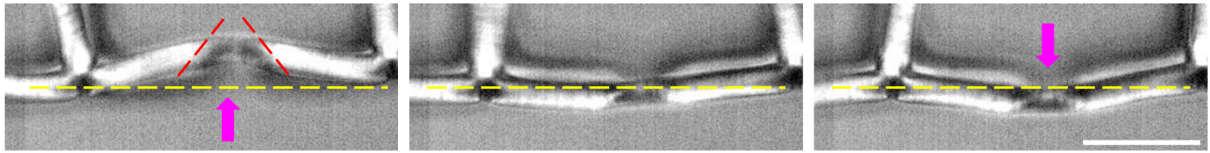

**Fig. S6| Asymmetric oscillation of an acoustic-deformable microbeam with a wedge-shaped soft hinge.** Left: a moment of upward oscillation; Mid: horizontal initial position; Right: a moment of downward oscillation. The amplitude of 45 μm of the upward oscillation is 2 times higher than that of the downward oscillation. The pink arrows denote the oscillation direction. The red dotted lines denote the wedge-shaped hinge. The yellow dotted lines denote the initial position. Scale bar 100 μm.

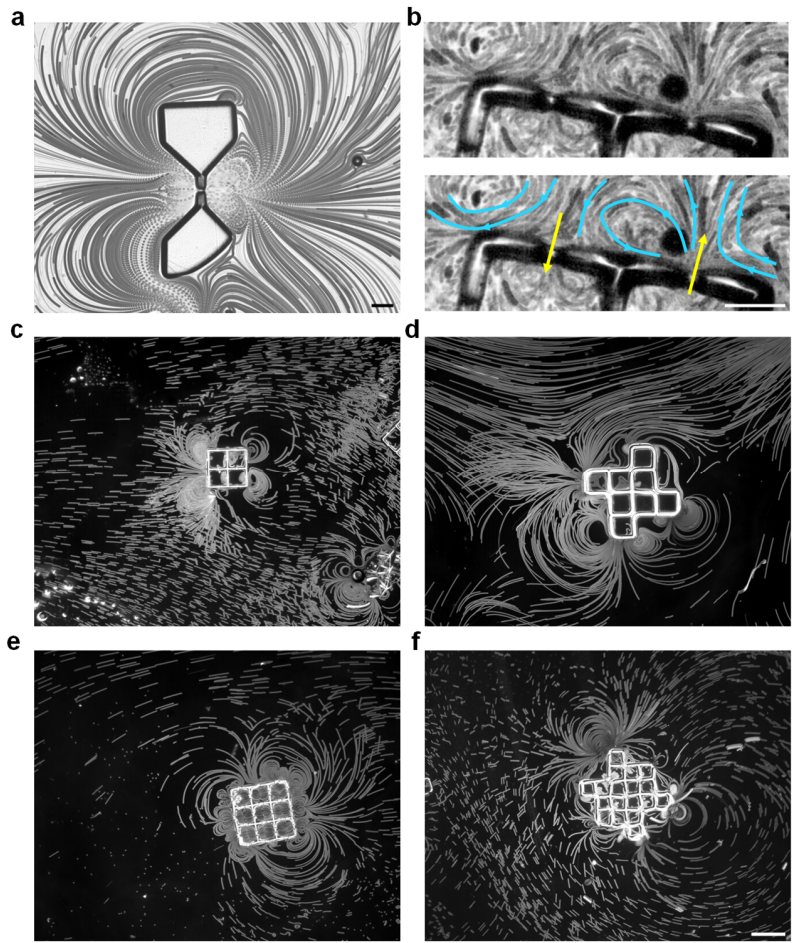

**Fig. S7| Acoustic streamlines around acoustic-transformable micromachines. a**, Asymmetric acoustic streaming around a single straight acoustic-deformable microbeam. Scale bar, 100 μm. **b**, Flow field generated by two oscillating soft hinges surrounding a T-shaped structure. Opposite flow directions indicate that the T-shaped center structure is subjected to forces in opposite directions. Scale bar, 50 μm. **c**, to **f**, Asymmetric acoustic streamlines around a four-square micromachine and a nine-square micromachine with the original shapes and their full-folded shapes. Scale bar, 200 μm. The tracer particle size is 10 μm. The acoustic excitation frequency and voltage are 6.5 kHz and 8 Vpp, respectively. The UV exposure time is 100 ms.

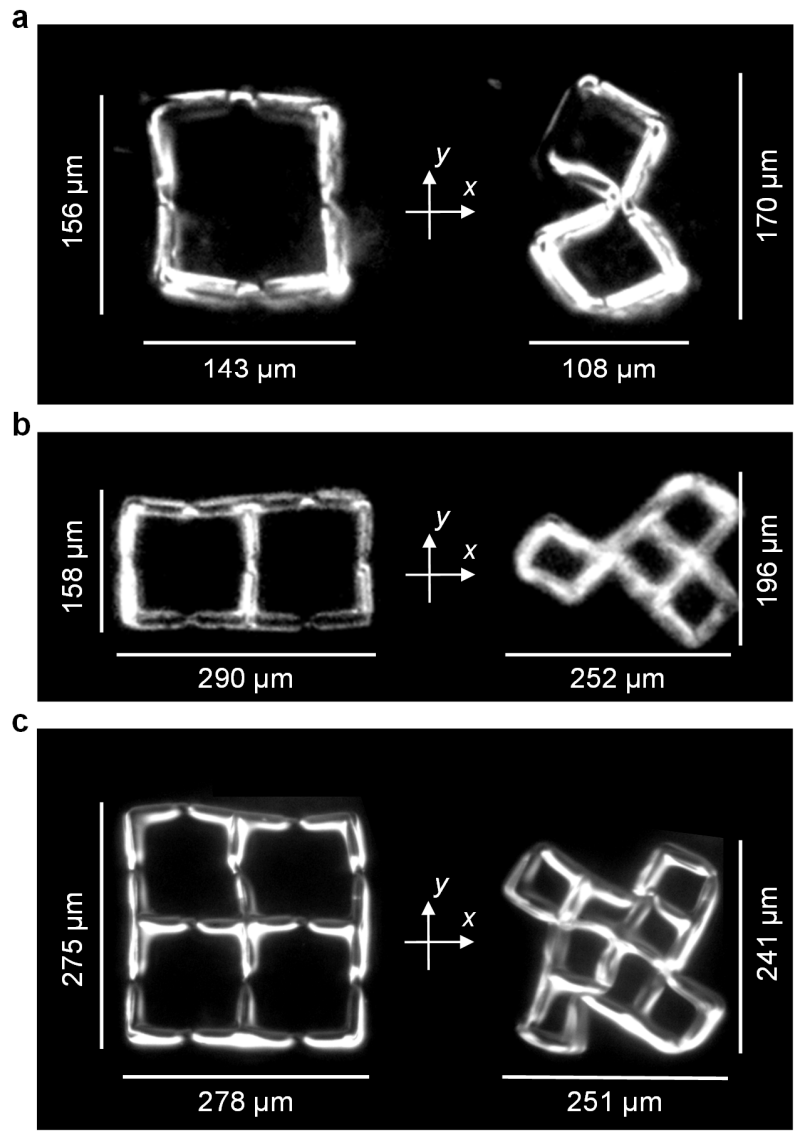

**Fig. S8| Calculation of the Poisson's ratio of acoustic-transformable micromachines. a**, A single square micromachines, **b**, a two-square micromachines, and **c**, a four square micromachine respectively show the Poisson's ratio of 2.73, 1.84, and -1.27.

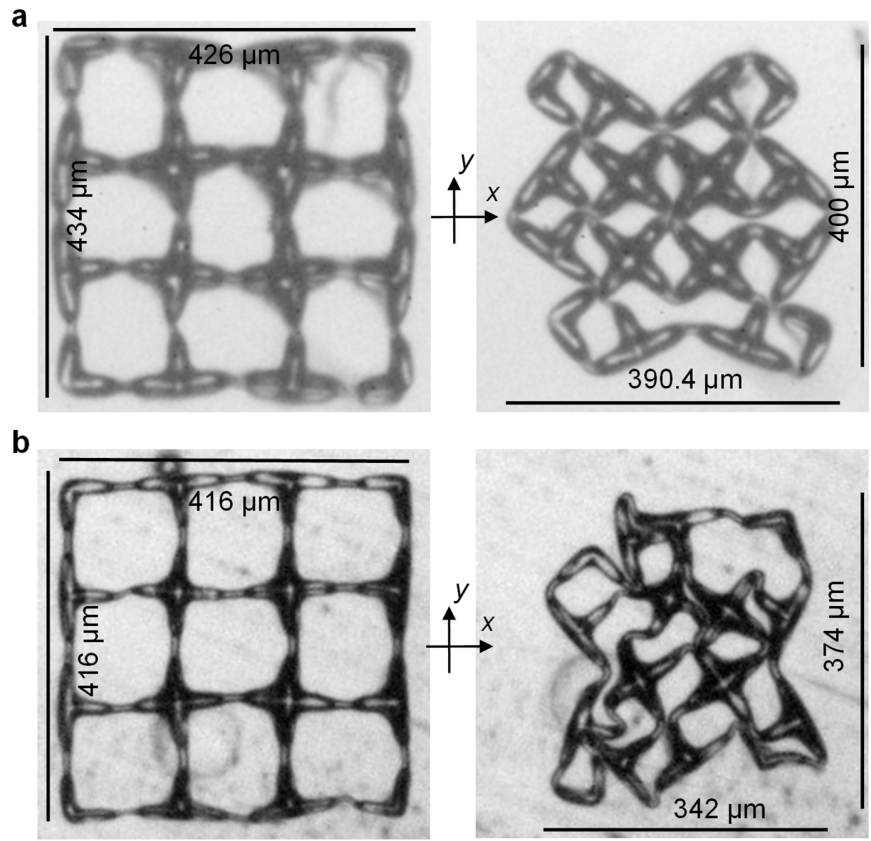

**Fig. S9| Calculation of the Poisson's ratio of nine-square acoustic-transformable micromachines. a**, A micromachine shows the Poisson's ratio of -1.07. The experimental result is consistent with the theoretical value of -1. **b**, A micromachine shows the Poisson's ratio of -1.76 due to the over-folded deformation.

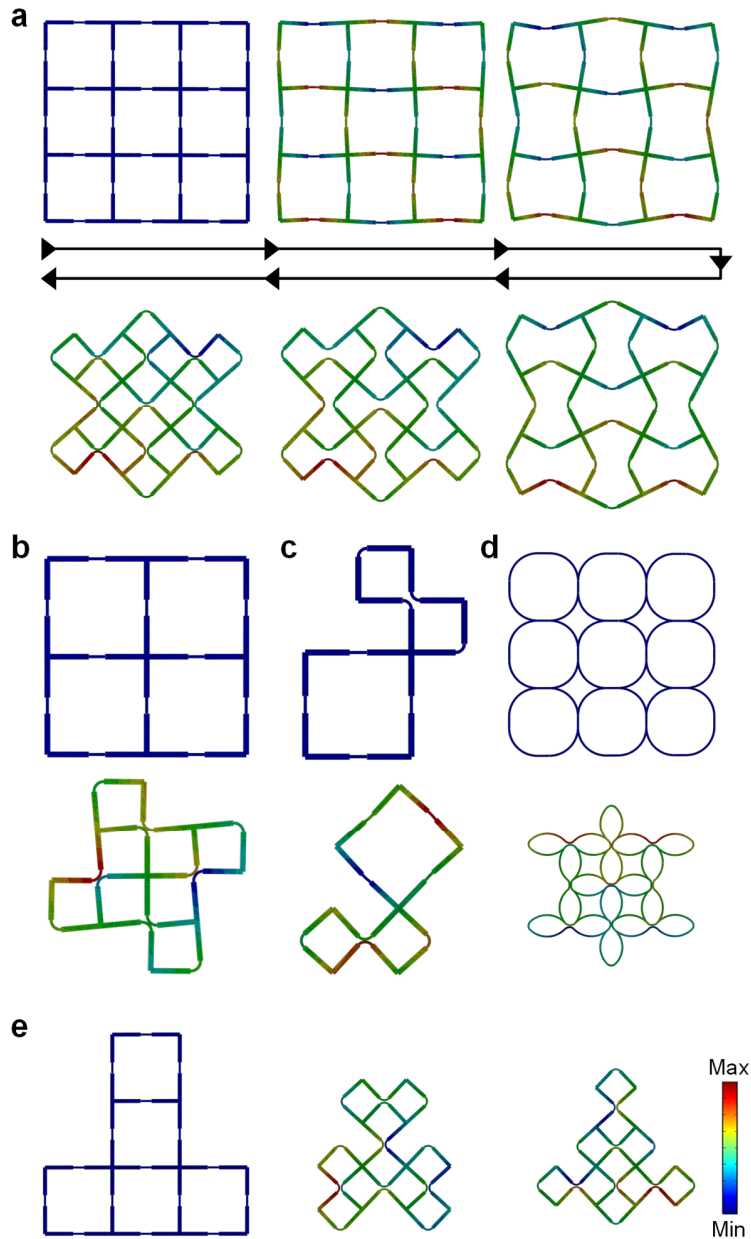

**Fig. S10| Numerical simulations of various acoustic-transformable micromachines. a**, Transformation sequence of a nine-square micromachine. **b**, to **e**, Original and transformed shapes of a four-square micromachine incorporating only straight acoustic-deformable microbeams, a push-pull micromachine incorporating both straight and curved microbeams, a nine-square micromachine with curved microbeams, and a T-shaped micromachine that can transform toward two isomeric shapes. The color bar denotes the displacement of elements. The finite element simulation model was developed with the commercial software, Abaqus, with the same dimensions as the experimental prototype. We applied pressure on the interconnected microbeams to demonstrate the possible transformation of each micromachine.

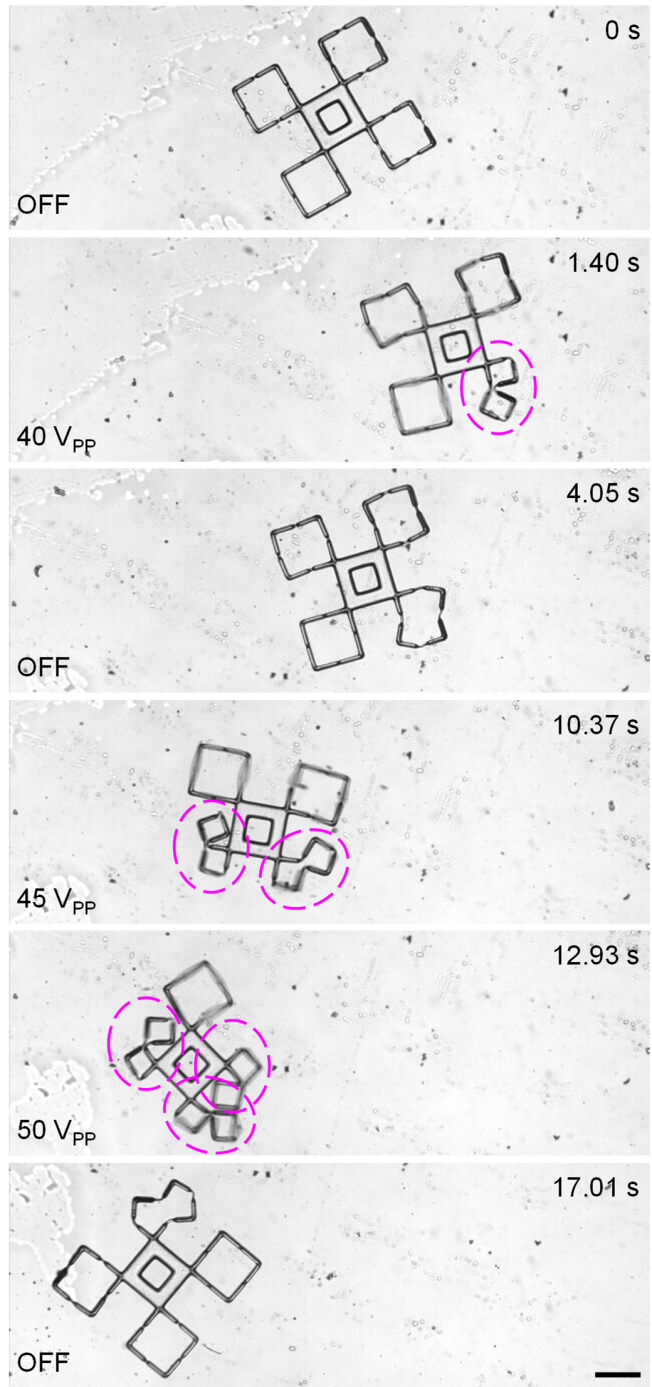

**Fig. S11| Dynamic selective shape transformation of a microrobot against acoustic excitation voltage.** The magenta dotted ellipse shows the folded leg. Scale bar, 100 μm.

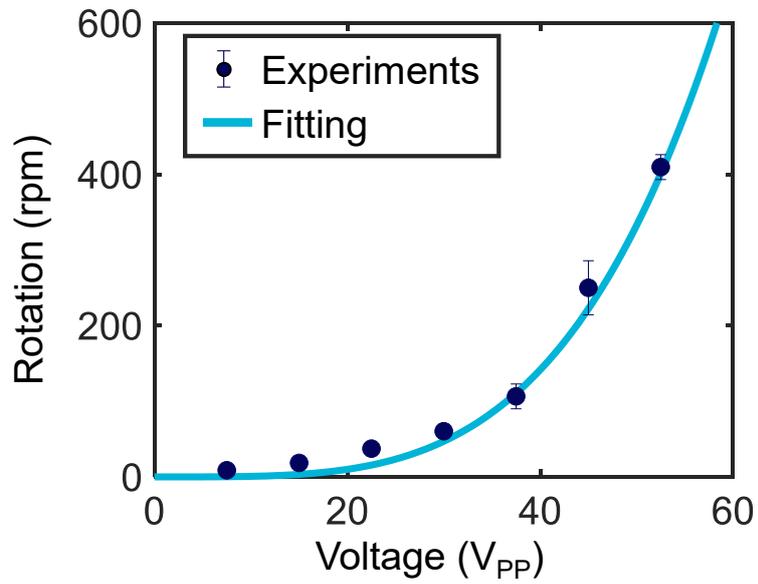

**Fig. S12| Microrobot rotational speed against acoustic excitation voltage.** The fitting is performed with a power function as $y = 0.0001043x^{3.828}$.

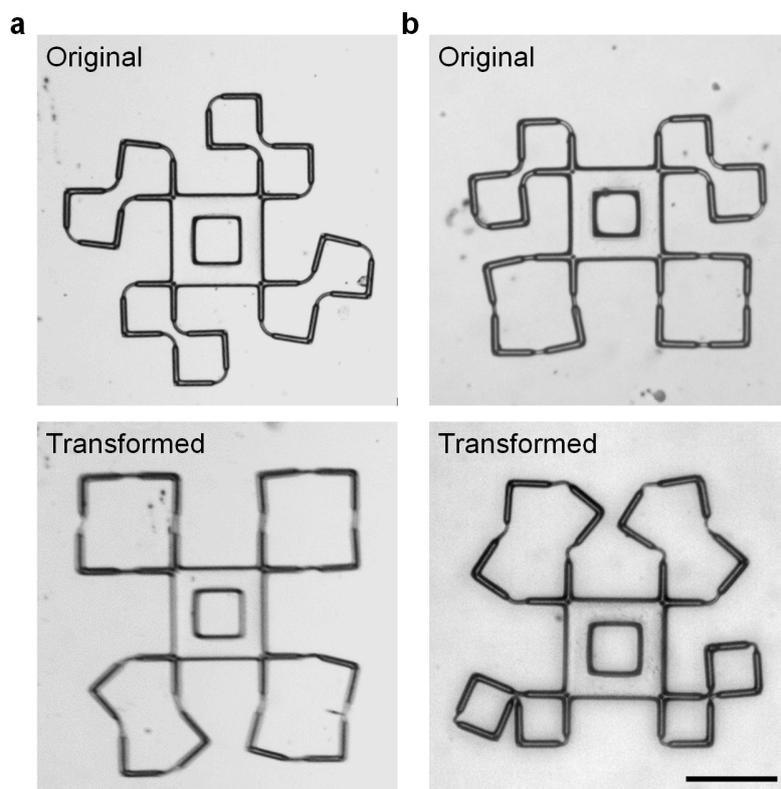

**Fig. S13| Acoustic-transformable microrobots. a**, A quadrupedal microrobot with four contracted legs, which can transform toward its full-extended shape in an acoustic field. **b**, A quadrupedal microrobot with two contracted legs and two extended legs, which can transform toward the mirror-transformed shape in an acoustic field. Scale bar, 100 μm.

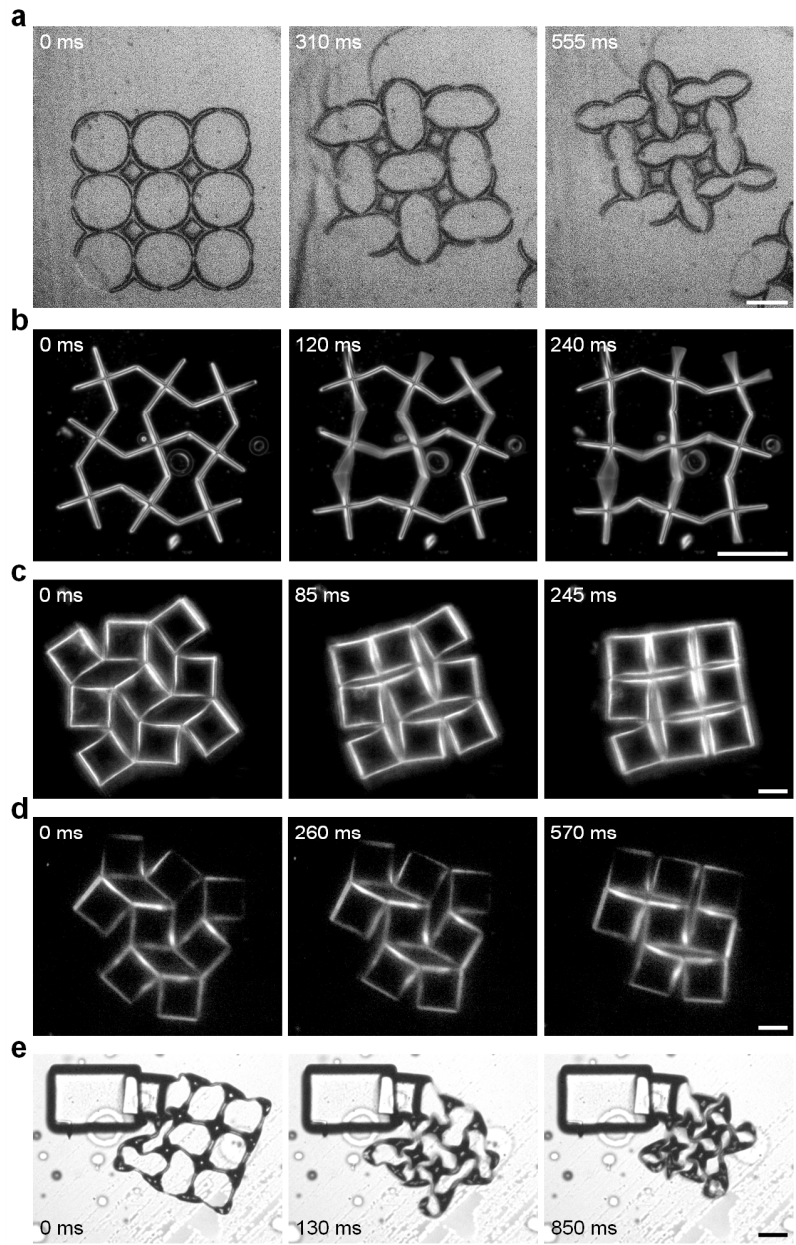

**Fig. S14| Acoustic-transformable micromachines with more geometric designs. a**, A micromachine with curved links shows folding in an acoustic field. **b**, A micromachine with point hinges shows extension in an acoustic field. **c**, A micromachine with block links and point hinges shows folding in an acoustic field. **d**, A micromachine shows folding in an acoustic field with a broken portion. **e**, A micromachine shows folding with an additional block. Scale bar 100 μm.

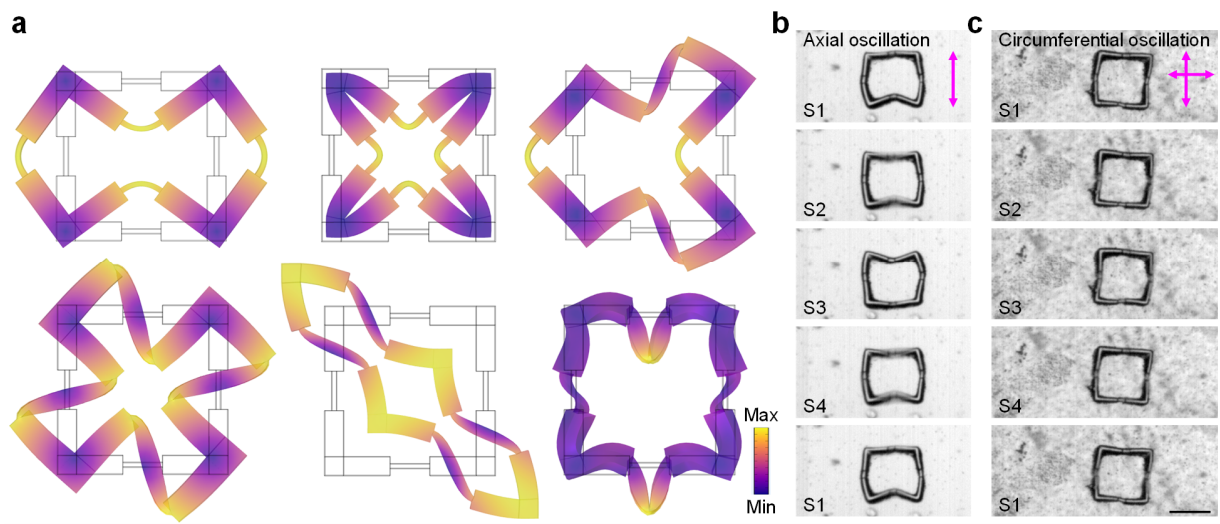

**Fig. S15| Different oscillation modes of a square micromachine. a**, Several simulated oscillation modes of the square micromachine. The color denotes mechanical stress. **b**, to **c**, Experimental observed axial oscillation and circumferential oscillation of the micromachine. The label shows the periodic oscillation phases. The acoustic excitation frequencies are 9.1 kHz and 9.2 kHz, respectively. Scale bar, 100 μm. The relationship between acoustic interaction and the oscillation modes of soft matter is out of the scope of this study and will be explored deeper in the future.

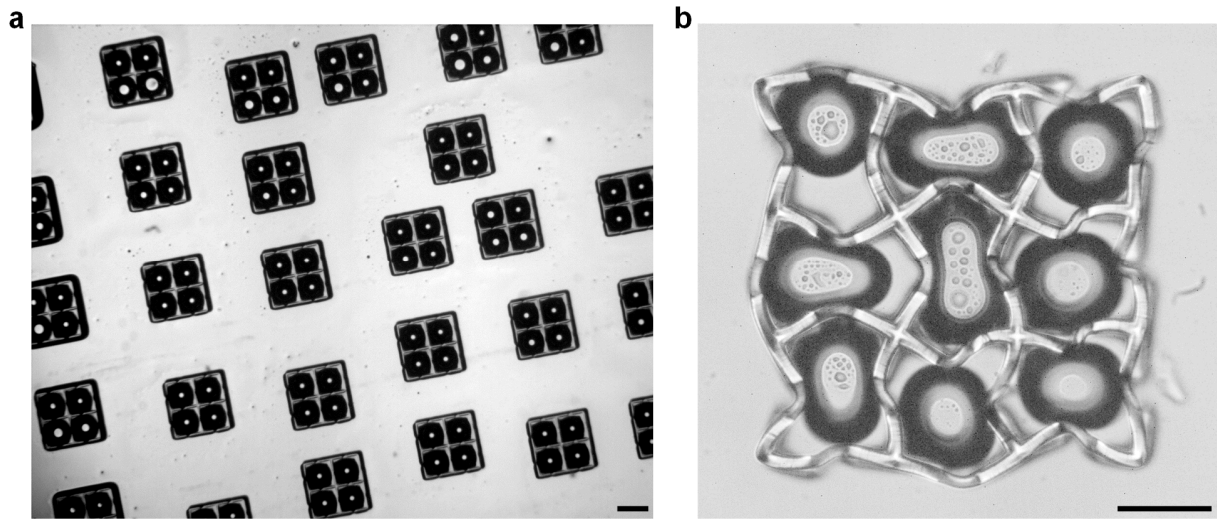

**Fig. S16| Bubbles are trapped by square micromachines. a**, Bubbles are simultaneously trapped by each sub-square of the micromachines. **b**, Bubbles are squeezed by the shape-shifting of the micromachine. Scale bar, 100 μm.

**Supplementary Table**

**Table 1. Optimization of nine-square acoustic-transformable micromachines.** Gray blocks and error symbols indicate that these micromachines cannot be deformed because the acoustic forces are too small; pink blocks and error symbols indicate that these micromachines are broken because the acoustic forces are too strong. The correct symbols indicate those machines capable of transformation. The UV exposure time and acoustic excitation frequency are 150 ms and 6.5 kHz.

| (*t*, *l*)(mm) | 30 $V_{PP}$ | 37.5 $V_{PP}$ | 45 $V_{PP}$ | 52.5 $V_{PP}$ | 60 $V_{PP}$ |
|---|---|---|---|---|---|
| (0.1, 0.1) | ✗ | ✓ | ✓ | ✗ | ✗ |
| (0.1, 0.3) | ✓ | ✓ | ✗ | ✗ | ✗ |
| (0.2, 0.1) | ✗ | ✗ | ✗ | ✓ | ✓ |
| (0.2, 0.3) | ✗ | ✗ | ✓ | ✓ | ✓ |
| (0.2, 0.5) | ✗ | ✓ | ✓ | ✓ | ✓ |
| (0.2, 0.9) | ✗ | ✗ | ✗ | ✓ | ✓ |
| (0.3, 0.1) | ✗ | ✗ | ✗ | ✗ | ✗ |
| (0.3, 0.3) | ✗ | ✗ | ✗ | ✗ | ✓ |

**Legends for Supplementary Videos**

**Video 1.**
**Ultrafast auxetic shape transformation showing by nine-square soft micromachines**. The acoustic excitation frequencies range from 5.5 kHz to 6.5 kHz and excitation voltages range from 30 $V_{PP}$ to 60 $V_{PP}$. To achieve this transformation behavior, an appropriate UV exposure time of 80 ms to 150 ms is necessary.

**Video 2.**
**Ultrafast selective shape transformation tuning by acoustic excitation voltages**. A quadrupedal microrobot can selectively fold its square-shaped legs with different hinge lengths under different acoustic excitation voltages. The leg with the longest soft hinge is first folded at a lower voltage.

**Video 3.**
**Dynamics of straight acoustic-deformable microbeams**. The oscillation and folding behavior of the straight acoustic-deformable microbeams under different boundary conditions are observed by a high-speed camera.

**Video 4.**
**Versatile Shape Transformation of Micromachines**. (i) Bilateral auxetic shape transformation. A four-square micromachine constructed from straight acoustic-deformable microbeams folds into its full-folded shape; while an inverse four-square micromachine constructed from curved acoustic-deformable microbeams is extended to its full-extended shape. (ii) Push-pull shape transformation. A micromachine constructed from both straight and curved acoustic-deformable microbeams shows the folded structure extends when the extended structure folds upon acoustic activation. (iii) Isomeric shape transformation. A T-shaped micromachine with the right wedge-shaped hinge transforms toward a "convex" shape; while the one with the left wedge-shaped hinge transforms toward a "concave" shape.

**Video 5.**
**Faster rotation of a quadrupedal microrobot**. Under a high acoustic excitation voltage, the microrobot transforms into its full-folded shape, causing a lower inertia moment and a faster rotational speed.

**Video 6.**
**Reliable and adaptable shape transformation**. Micromachines with more geometric designs also achieve ultrafast shape transformation.

**Supplementary References**